\documentclass[12pt,preprint]{aastex}
\usepackage{amsmath}
\begin{document}
\title{Time-Resolved and Energy-Resolved Polarizations of GRB Prompt Emission}
\author{Mi-Xiang Lan$^{1}$, and Zi-Gao Dai$^{2, 3}$}
\affil{$^{1}$Center for Theoretical Physics and College of Physics, Jilin University, Changchun, 130012, China; lanmixiang@jlu.edu.cn \\
$^{2}$School of Astronomy and Space Science, Nanjing University, Nanjing 210093, China; dzg@nju.edu.cn \\
$^{3}$Key Laboratory of Modern Astronomy and Astrophysics (Nanjing University), Ministry of Education, China \\}

\begin{abstract}
Besides light curves and spectra, polarization provides a different powerful tool of studying the $\gamma-$ray burst (GRB) prompt phase. Compared with the time-integrated and energy-integrated polarization, time-resolved and energy-resolved polarization can deliver more physical information about the emitting region. Here we investigate time-resolved and energy-resolved polarization of GRB prompt emission using the synchrotron models. We find that the equal arrival time surface effect is very important in shaping the PD curves when the physical conditions of emitting region changes violently with radius. Polarization properties are neither correlated with the spectral lag nor the peak energy evolution patterns. Polarization properties with a mixed magnetic field are very similar to those for a corresponding ordered magnetic field but the former has a smaller polarization degree. The emission at the MeV peak can be highly polarized for a synchrotron model while it is unpolarized as predicted by a dissipative photosphere model. Future energy-resolved polarization observations can distinguish between these two models.
\end{abstract}


\section{Introduction}

Gamma-ray bursts (GRBs) are intensive $\gamma$-ray transients occurring at the cosmological distances. After more than two decades studies (since 1997), their emission mechanism remains mysterious. Light curves of GRB prompt emission usually show rich diversity. Different from the variety of GRB prompt light curves, their spectra are often described by an empirical relation, i.e., the Band function (Band et al. 1993), which is a two-segment power-law smoothly connected at a peak energy ($E_{p,obs}$). The origin of the Band function is also a mystery. The low-energy photon spectral indexes of the Band spectra for long GRBs are usually around $-1$, which is inconsistent with the prediction of synchrotron emission. Recently, a series of studies had been made to interpret the observations of GRB prompt phase, including the synchrotron origin of the Band spectrum (Uhm \& Zhang 2014), the spectral lag (Uhm \& Zhang 2016) and the two peak energy evolution patterns (Uhm, Zhang \& Racusin 2018).

The physical mechanism of the GRB prompt phase has been poorly known despite two decades of studies. Polarization is very sensitive to radiation mechanisms, jet structure and observational geometry, which in turn can be used as a powerful tool of investigating these properties. Recently, the POLAR team published their polarization measurements of 5 GRBs (Zhang et al. 2019). Then a detailed time-resolved polarization analysis was performed by Burgess et al. (2019). However, theoretical studies of time-resolved and energy-resolved polarization in the GRB prompt phase are very rare. In the following, we briefly review the theoretical progress made in the GRB prompt phase.

To interpret the large polarization degree (PD) observations of GRB 021206 (Coburn \& Boggs 2003), Lyutikov et al. (2003) derived the polarization of an electromagnetically dominated spherical outflow and a large-scale ordered magnetic field, which is carried out from the central engine and assumed to be latitude circles in the spherical shell. If the aperture of the line of sight (LOS) from the polar axis of the spherical shell exceeds the $1/\Gamma$ cone\footnote{$\Gamma$ is the bulk Lorentz factor of the spherical shell.}, then because the symmetric axis is outside the observational cone and circles of magnetic line are incomplete in $1/\Gamma$ cone, the resultant PD approaches a maximum value. Using the observations of GRB 021206, Granot (2003) tested three polarization models with different magnetic field configurations (MFCs). They found that the aligned magnetic field is most powerful in interpreting the high PD observations.

The polarization in both GRB prompt and afterglow phase was reviewed in Lazzati (2006). Because of relativistic motion of GRB ejecta, only a small fraction of the ejecta (i.e., $1/\Gamma$ cone) can be seen in the observer frame. This has two effects on the polarization. One is an increase of regularity of a magnetic field in the observational cone, leading to a larger observed PD. The other is that the boosting direction at the edge of $1/\Gamma$ cone has an angle with that of LOS, which results in rotation of the local position angle (PA) at the edge and hence a reduction of the observed PD. Polarization models are divided into two classes, intrinsic and geometric. The intrinsic model involves the synchrotron emission, while the geometric model corresponds to a special observational geometry of a narrow jet. The asymmetry of the intrinsic model is provided by the magnetic field but offered by observational geometry for the geometric model.

Polarization of GRB prompt emission was considered by assuming the Band spectrum by Toma et al. (2009). In their model, synchrotron emission was assumed and the time-integrated Stokes parameters, describing linear polarization, were constructed. They discussed rich contents, including  polarization properties of synchrotron emission with a large-scale ordered toroidal magnetic field confined in the jet surface and with a two-dimensional random magnetic field. And also the Compton Drag mechanism was discussed. PDs of synchrotron emission with a toroidal magnetic field are large when the LOS is within the jet cone but not aligned with the jet axis. For synchrotron emission with a two-dimensional random magnetic field or the Compton Drag model, a large PD is obtained only for off-axis observation. In deriving the local PD $\pi_0=(\tilde{\alpha}+1)/(\tilde{\alpha}+5/3)$ of a jet element, relativistic electrons are assumed to be slow cooling and the power-law index $p$ of relativistic electrons is related to the photon spectral index $\tilde{\alpha}$ by $p=2\tilde{\alpha}+1$. This is inconsistent with the fast cooling in the GRB prompt phase inferred from observations.

Gill, Granot \& Kumar (2019) discussed the polarization of GRB prompt phase by considering different mechanisms (including both intrinsic and geometric models) with various jet structures. Literally, polarization in the various photosphere models were discussed extensively. Lundman et al. (2014) studied the polarization properties of photospheric emission of a top-hat jet with a surrounding shear layer. Polarization of photospheric emission with a stratified jet were discussed by Ito et al. (2014). Polarization in the dissipative photosphere model was calculated and this model predicted that if the jet is uniform at an angular range of $\theta>1/\Gamma$, then the MeV peak and the above energy-band are unpolarized (Lundman et al. 2018).

In this paper, we investigate time-resolved and energy-resolved polarization during the GRB prompt phase by using the emission model proposed by Uhm \& Zhang (2015, 2016) and Uhm, Zhang \& Racusin (2018). The paper is arranged as follows. In Section 2, we give out our polarization models. In Section 3, we numerically explore the polarization properties of GRB prompt emission using models described in Section 2. Finally,  conclusions and discussion are exhibited in Section 4.

\section{The Models}

Following some previous studies (Uhm \& Zhang 2015, 2016; Uhm, Zhang \& Racusin 2018), we consider a relativistic jet shell expanding radially. The electrons are injected into the shell at an isotropic rate $R_{inj}$ and radiate synchrotron emission. The spectral power of the emission may be depicted by the Band function on average and it reads
\begin{equation}
P'_{\nu'}(\nu')=P'_0H_{en}(x),
\end{equation}
where $P'_0$ represents the spectral power magnitude of a single electron and $H_{en}(x)$ describe the shape of the emission spectrum with $x=\nu'/\nu'_{ch}$. $\nu'$ and $\nu'_{ch}$ are the observational and critical frequencies in the fluid co-moving frame, respectively. For an electron with mass $m_e$ and charge $q_e$, we have (Rybicki \& Lightman 1979)
\begin{equation}
P'_0=\frac{3\sqrt{3}}{32}\frac{m_ec^2\sigma_TB'}{q_e},\ \ \ \ \nu'_{ch}=\frac{3}{16}\frac{q_eB'}{m_ec\gamma_{ch}^2},
\end{equation}
where $c$ is the speed of light and $B'$ is the strength of magnetic field in the co-moving frame. With these assumptions, all the electrons are on average have the same Lorentz factor $\gamma_{ch}$.
\begin{equation}
H_{en}(x)=\begin{cases}
x^{\alpha_B+1}\exp(-x), & \text{$x\leq x_c$}, \\ x_c^{x_c}\exp(-x_c)x^{\beta_B+1}, & \text{$x\geq x_c$},
\end{cases}
\end{equation}
where $x_c=\alpha_B-\beta_B$. It is assumed that the shell begins to emit photons at radius $r_{on}$ at a burst source time $t_{on}$. Thus the photons emitted at radius $r$ at a burst source time $t$ arrive at the observer at an observer time $t_{obs}$ (Uhm \& Zhang 2016)
\begin{equation}
t_{obs}=\left[t-\frac{r}{c}\cos\theta-t_{on}+\frac{r_{on}}{c}\right](1+z).
\end{equation}
With the spectral power of single electrons, the observed spectral flux $F_{\nu}$ of a GRB at redshift $z$ can be constructed (Uhm \& Zhang 2015). The integration is performed on the equal arrival time surface (EATS).
\begin{equation}
F_{\nu}=\frac{1+z}{4\pi D_L^2}\int\frac{\mathcal{D}^2}{\Gamma}\frac{c}{4\pi r}NP'_0H_{en}(x)(\sin\theta'_B)^{\tilde{\alpha}+1}d\phi dt,
\end{equation}
where $D_L$ is the luminosity distance of the source, $\mathcal{D}=1/\Gamma(1-\beta\cos\theta)$ is the Doppler factor and $N=\int R_{inj}dt/\Gamma$ is the isotropic total electron number in the shell. $\nu$ is the observational frequency and $\nu'=\nu(1+z)/\mathcal{D}$. $\Gamma$ and $\beta$ are the bulk Lorentz factor and dimensionless velocity of the shell. $\phi$ is the angle in the plane of sky between the projection of jet axis and the projection of the radial direction of a local fluid element. The spectral index $\tilde{\alpha}$ equals to $\alpha_B$ when $x\leq x_c$ and is $\beta_B$ when $x\geq x_c$. Then the Stokes parameters, describing linear polarization, $Q_{\nu}$ and $U_{\nu}$ will be expressed as
\begin{equation}
Q_{\nu}=\frac{1+z}{4\pi D_L^2}\int\frac{\mathcal{D}^2}{\Gamma}\frac{c}{4\pi r}NP'_0H_{en}(x)(\sin\theta'_B)^{\tilde{\alpha}+1}\Pi_p\cos2\chi_pd\phi dt,
\end{equation}
and
\begin{equation}
U_{\nu}=\frac{1+z}{4\pi D_L^2}\int\frac{\mathcal{D}^2}{\Gamma}\frac{c}{4\pi r}NP'_0H_{en}(x)(\sin\theta'_B)^{\tilde{\alpha}+1}\Pi_p\sin2\chi_pd\phi dt,
\end{equation}
where $\Pi_p$ and $\chi_p$ are the local PD and PA.

The above expressions of three Stokes parameters (i.e., $F_{\nu}$, $Q_{\nu}$ and $U_{\nu}$) seem different from that exhibited before (Lan, Wu \& Dai 2016a). In fact, they can be transformed to similar form by replacing the integration variable $t$ with $\theta$. On an EATS, we take a differential and get
\begin{equation}
dt=\frac{r\sin\theta d\theta}{(1-\beta\cos\theta)c}.
\end{equation}
Replacing Eq. (8) in Eqs. (5)-(7), similar formulas are obtained
\begin{equation}
F_{\nu}=\frac{1+z}{4\pi D_L^2}\int\mathcal{D}^3\sin\theta d\theta\int d\phi\frac{NP'_0H_{en}(x)(\sin\theta'_B)^{\tilde{\alpha}+1}}{4\pi},
\end{equation}
\begin{equation}
Q_{\nu}=\frac{1+z}{4\pi D_L^2}\int\mathcal{D}^3\sin\theta d\theta\int d\phi\Pi_p\cos2\chi_p\frac{NP'_0H_{en}(x)(\sin\theta'_B)^{\tilde{\alpha}+1}}{4\pi},
\end{equation}
and
\begin{equation}
U_{\nu}=\frac{1+z}{4\pi D_L^2}\int\mathcal{D}^3\sin\theta d\theta\int d\phi\Pi_p\sin2\chi_p\frac{NP'_0H_{en}(x)(\sin\theta'_B)^{\tilde{\alpha}+1}}{4\pi}.
\end{equation}

If both Stokes parameters $Q_\nu$ and $U_\nu$ of the jet emission are non-zero, PD ($\Pi$) and PA ($\chi$) of the GRB prompt phase can be expressed by
\begin{equation}
\Pi=\frac{\sqrt{Q^2_\nu+U^2_\nu}}{F_\nu},
\end{equation}
and
\begin{equation}
\chi=\frac{1}{2}\arctan\left(\frac{U_\nu}{Q_\nu}\right).
\end{equation}
The above formula for the PA $\chi$ is incomplete, when $Q_\nu>0$, the final PA equals to $\chi$; when $Q_\nu<0$, if $U_\nu>0$ then the final PA is $\chi+\pi/2$, if $U_\nu<0$ then the final PA is $\chi-\pi/2$ (Lan, Wu \& Dai 2018).

Literally, for polarization of synchrotron emission, three classes of MFCs are discussed, large-scale ordered, mixed (including both ordered and random parts) and random (Sari 1999; Granot \& K\"{o}nigl 2003; Lyutikov 2003; Toma et al. 2009; Lan et al. 2019). And also three kinds of large-scale ordered magnetic fields had been discussed, aligned, toroidal and radial (see Fig. 1 of Lan et al. (2019) for a view of these three MFCs). PDs of the synchrotron emission with an istropic 3-dimensional random magnetic field are always zero, independent of parameters (Lan et al. 2019). Here, for the emitting jet shell discussed in this paper, the PD of the 2-dimensional random magnetic field will roughly be 0 if the LOS is well inside the jet cone (Toma et al. 2009). For the toroidal magnetic field, its polarization properties are very similar to that of an aligned one if the LOS is neither close to the jet axis nor far outside the jet cone (Lan et al. in preparation). Therefore, we only consider the large-scale ordered aligned magnetic field and the mixed magnetic field with an aligned ordered part (Lan et al. 2019).

For the synchrotron emission with an ordered aligned magnetic field, local PA and the pitch angle can be found in Eqs. (26) and (25) of Lan, Wu \& Dai (2016a). In this paper, the single-energy electrons are assumed and hence $\Pi_p=\Pi_0$ with $\Pi_0=G(x)/F(x)$, where $G(x)=xk_{\frac{2}{3}}(x)$ with the modified Bessel function of $\frac{2}{3}$ order $k_{\frac{2}{3}}(x)$ and $F(x)$ is the dimensionless synchrotron spectrum of a single electron. It can be found that $\Pi_0$ approaches $0.5$ when $x$ approaches 0, it approaches 1 when $x$ approaches infinity and its value increases quickly at vicinity of $x\sim1$. For the synchrotron emission in a mixed magnetic field with an aligned ordered part, local PD and PA are exhibited in Eqs. (4) and (5), with $\Pi_0$ given by the above expression for the single-energy electrons.

\section{Numerical Results}

Only a small fraction of GRBs prompt spectra can be modeled by the fast cooling synchrotron model (Zheng et al. 2012; Oganesyan et al. 2019). This fast cooling problem of GRB prompt phase disfavors the small-radius internal shock model (Ghisellini et al. 2000). This is the motivation of large-radius synchrotron models, such as the ICMART model (Zhang \& Yan 2011) and large-radius internal shock model (Hasco\"{e}t et al. 2012). Considering a decaying magnetic field, the fast cooling synchrotron spectra can have a Band-function-like form (Uhm \& Zhang 2014) and such a model can fit the GRB prompt spectra well (Zhang et al. 2016).

In the GRB prompt phase, two peak energy evolution patterns have been found, i.e., hard-to-soft and intensity tracking (Liang \& Kargatis 1996; Ford et al. 1995; Kaneko et al. 2006; Lu et al. 2010). And also two spectral lag were discovered (Norris et al. 1996, 2000; Liang et al. 2006). All these observational properties were reproduced by models (Uhm \& Zhang 2016; Uhm, Zhang \& Racusin 2018). Here, we take 6 models, i.e., [2$b_i$], [2$c_i$], [2$d_i$], [2$b_m$], [2$c_m$] and [2$d_m$], in Uhm, Zhang \& Racusin (2018) to discuss their polarization properties. For models of [2$b_i$], [2$c_i$] and [2$d_i$], the hard-to-soft $E_{p,obs}$ evolution pattern is recovered and the spectral lags are all positive. In models [2$b_m$], [2$c_m$] and [2$d_m$], $E_{p,obs}$ evolutions exhibit a intensity tracking pattern. The light curves for [2$b_m$] and [2$c_m$] show positive spectral lags, while it shows negative spectral lag for [2$d_m$] model. The only difference for the models with subscribe ``m" and ``i" is their $\gamma_{ch}$ profile.

Following Uhm, Zhang \& Racusin (2018), a jet, which begins to radiate at radius $r_{on}$ and cease at radius $r_{off}$, undergos bulk acceleration. The bulk Lorentz factor is assumed to be a power-law form with radius. The half-opening angle of the jet is taken as $\theta_j=0.1$ rad, which is larger than the maximum angular range given by the minimum radius $r_{on}$ on EATS.
\begin{equation}
\Gamma(r)=\Gamma_0(r/r_0)^s.
\end{equation}
The magnetic field strength of the emitting region in the co-moving frame decays with radius (Uhm \& Zhang 2014),
\begin{equation}
B(r)=B_0(r/r_0)^{-b}.
\end{equation}
The evolutions of $\gamma_{ch}$ for ``i" and ``m" models are described by the following functions (Uhm \& Zhang 2014). It is assumed to be a single power-law with radius for models with subscribe ``i'',
\begin{equation}
\gamma_{ch}(r)=\gamma_{ch}^0(r/r_0)^g,
\end{equation}
where we take $\gamma_{ch}^0=5\times10^4$ and $g=-0.2$ for ``i" models. The profile of electron Lorentz factor is a broken power-law in radius for ``m" models, which reads
\begin{equation}
\gamma_{ch}(r)=\gamma_{ch}^m\times\begin{cases}
(r/r_m)^g, & \text{$r\leq r_m$}, \\ (r/r_m)^{-g}, & \text{$r\geq r_m$},
\end{cases}
\end{equation}
where the normalization is taken at $r_m=2\times10^{15}$ cm and $\gamma_{ch}^m=2\times10^5$. The power-law index is taken as $g=1.0$.

Model parameters for these 6 models we take are the same as those in Uhm, Zhang \& Racusin (2018), i.e., $\alpha_B=-0.8$, $\beta_B=-2.3$, $R_{inj}=10^{47}\ s^{-1}$, $r_0=10^{15}$ cm, $\Gamma_0=250$, $s=0.35$, $r_{on}=10^{14}$ cm, $r_{off}=3\times10^{16}$ cm and $B_0=30$ G. The power-law index of the magnetic field strength $b$ is 1.0 for [2$b_i$] and [2$b_m$], is 1.25 for [2$c_i$] and [2$c_m$], and is 1.5 for [2$d_i$] and [2$d_m$]. The orientation of the purely aligned magnetic field or the ordered part of the mixed magnetic field is assumed to be $\delta=\pi/6$. For the mixed magnetic field, the magnetic field strength ratio of the ordered part to the random part is taken as $\xi_B=1.1$.

With the models described in Section 2, we numerically calculate the polarization evolution for both ``i" and ``m" models. The polarization behavior of a toroidal MFC is very similar to that of an aligned one, if the viewing angle is inside the jet cone but not exactly zero (Lan et al. in preparation). For a 2-dimensional random magnetic field, its PD is usually very small, when the viewing angle is inside the jet cone (Toma et al. 2009). Therefore, it is not considered here. The PD is always 0 for a 3-dimensional isotropic random magnetic field because of its symmetry (Lan et al. 2019). Finally, here, we only consider two kinds of MFCs, i.e., a large-scale ordered aligned MFC (Granot \& K\"{o}nigl 2003) and a mixed MFC with an aligned ordered part (Lan et al. 2019).

To interpret our following results, by considering [2$b_i$] model as an example, we calculate $PD(\theta)$, i.e., local PD of emission from the electrons within $\theta$ circle, we find when $\theta$ circle is outside the local $1/\Gamma$ cone (i.e., $\theta\Gamma>1$), $PD(\theta)$ is about $10^{-5\sim-2}$. When $\theta\sim1/\Gamma$, $PD(\theta)$ is at a magnitude of $10^{-2}$. Then when $\theta$ circle is inside $1/\Gamma$ cone (i.e., $\theta\Gamma<1$), $PD(\theta)$ is at a level of $10^{-1}$ and will increase quickly with a decreasing $\theta\Gamma$\footnote{Through our calculation, we find that $\theta\Gamma(r)$ will decrease with $r$.}. Finally, we conclude that local $PD(\theta)$ is large when $\theta\Gamma<1$ and is very tiny when $\theta\Gamma>1$.

In our calculation, EATS effect is considered. On one EATS, with increasing of radius $r$, $\theta$ decreases and $\Gamma(r)$ increases, so the Doppler boost factor increases. The Doppler factor is larger for a large radius than that at a small radius, but the magnetic field becomes weaker at a larger radius. So the ratio $\tilde{f}(t_{obs})$ of the flux from $\theta$ circles with $\theta\Gamma(r)<1$ and with $\theta\Gamma(r)>1$ is indefinite.
\begin{equation}
\tilde{f}(t_{obs})=\frac{\int_{r_c}^{r_{max}}dF_{\nu}}{\int_{r_{on}}^{r_c}dF_{\nu}},
\end{equation}
where $r_{max}=\min(r_{off},r(t_{obs},\theta=0))$\footnote{$r(t_{obs},\theta=0)$ is the maximum radius on the EATS with observational time $t_{obs}$.} and $\theta\Gamma(r_c)=1$.
Therefore, PD of the jet increases with $\tilde{f}(t_{obs})$ because the proportion of highly-polarized low-latitude emission increases.

Fig. 1 shows our results for a large-scale ordered aligned MFC with ``i" models. The spectral lags are all positive and the peak energy $E_p$ exhibits a hard-to-soft evolution pattern. PDs decay for all three ``i" models and for all the energy bands calculated. At any observational time, PDs of the [2$b_i$] model is largest, then the [2$c_i$] model, finally the [2$d_i$] model. The only difference of three ``i" models is that the decaying indice of the magnetic field $b$ varies.

We consider models [2$b_i$] and [2$d_i$] as an example to illuminate that PDs of [2$d_i$] model is smaller than that of [2$b_i$] model. Taking $t_{obs}=1$ s and $\nu_{obs}=1$ MeV, we have $\tilde{f}=2.4$ for [2$b_i$] model, while $\tilde{f}=0.7$ for [2$d_i$] model. Combining our conclusion about the jet PD and its connection with $\tilde{f}(t_{obs})$ value, PDs of [2$b_i$] model is larger than that of [2$d_i$] model. We take $\nu_{obs}=1$ MeV with [2$b_i$] model as an example to explain that PDs will decrease with time. When $t_{obs}=$0.2 s, 0.4 s, 1.0 s, 2.1 s, 3.0 s and 4.0 s, the corresponding $\tilde{f}(t_{obs})$ values are $\infty$, 18, 2.4, 1.1, 0.86 and 0.69. Then a decaying PD is obvious.

Comparing three ``i" models at same observational time and at same energy band, PD decreases from [2$b_i$] to [2$d_i$]. If the magnetic field strength is moderate (i.e., not strong enough to obviously suppress radiation), the radiation is brightened with an increase of the magnetic field strength. On an EATS, if the magnetic field strength decays faster, the proportion $\tilde{f}(t_{obs})$ becomes smaller, so does the PD. And also PD decays faster from [2$b_i$] to [2$d_i$], which is also attributed to a steeper decaying $\tilde{f}(t_{obs})$ because of the steeper-decaying magnetic field strength. After $t_{obs}=4$ s, PD curves of all the models decay quickly to 0. At $t_{obs}=4$ s, the maximum radius on its EATS reaches $r_{off}$. Notice that the emission is assumed to turn off suddenly at $r_{off}$, when $t_{obs}>4$ s, there is a non-emitting region in the $1/\Gamma$ cone and the area of this non-emitting region increases with $t_{obs}$. Then the proportion $\tilde{f}(t_{obs})$ decreases quickly and so does the jet PD.

PAs of all three ``i" models keep to be constant at most times of the burst. There are abrupt $90^\circ$ PA changes at late times of some PA curves, where the flux is so low that makes the polarization measurement become very difficult.

Fig. 2 shows our results for the mixed magnetic field with an aligned ordered part for ``i" models. The profiles of the PD curves with such a mixed magnetic field are very similar to the corresponding PD curves with a purely aligned magnetic field, only with smaller PD values. PAs of ``i" models with such a mixed MFC are all constant during the burst for all calculated wavebands. Polarization properties of a mixed magnetic field with an aligned ordered part and a purely aligned magnetic field are alike. This conclusion is consistent with our initial setup for the mixed magnetic field (Lan et al. 2019). When $\xi_B=1.1$, the polarization flux of such a mixed magnetic field is determined by its ordered part, while the total flux, including the unpolarized part, is generated under the total magnetic field. The role of a 3-dimensional random magnetic field is the enhancement of the total flux and then the suppression of the final PD.

Polarized curves for three ``m" models with a purely aligned ordered magnetic field are exhibited in Fig. 3. The $E_{p,obs}$ evolution pattern is intensity-tracking. The spectral lag is positive for models [2$b_m$] and [2$c_m$] but it is negative for model [2$d_m$]. For all three ``m" models, PD curves at 1 MeV and 300 keV decay all the way, while there are small bumps around 2 s for 100 keV and 30 keV PD curves. We then do a set of numerical experiments and take [2$b_m$] as an example to illuminate. When $\nu_{obs}=$ 1 MeV or 300 keV, the parameter $\tilde{f}(t_{obs})$ always decays, which means that the proportion of the highly polarized emission decreases and then leads to a decreasing PD. But for $\nu_{obs}=$ 100 keV or 30 keV, we find there is a small bump of $\tilde{f}(t_{obs})$ and the position of this bump is slightly before the position of the PD bump. An increase of $\tilde{f}(t_{obs})$ roughly indicates an increasing PD value. PAs of three ``m" models keep to be constant during the main burst. Fig. 4 shows PD and PA evolutions of a mixed magnetic field with an aligned ordered part. Again the polarization properties of a mixed magnetic field are very similar to that of a corresponding purely ordered magnetic field. Depending on $\xi_B$ parameter, PDs of a mixed magnetic field will range from 0 to the value in a corresponding ordered magnetic field.

Comparing ``i" and ``m" models, before $t_{obs}=1$ s, PDs of ``m" model are larger than those of the corresponding ``i" model. The only difference of these two models is the evolution of $\gamma_{ch}$, which results in the difference of parameter $\tilde{f}(t_{obs})$ and local PD $\Pi_0$. Because $\gamma_{ch}$ of ``m" model is smaller than that of ``i" model when $r<5.6\times10^{14}$ cm, $x$ value of ``m" model is larger than that of ``i" model. Taking $\nu_{obs}=1$ MeV and $t_{obs}=0.5$ s of [2$b_i$] and [2$b_m$] models as two examples, we then calculate the $x/x_c$ value and find that it is always larger than unity for these two examples. Because the decay slope of $\gamma_{ch}$ with ``i" model is very flat ($g=-0.2$), changing range of $x/x_c$ value with radius $r$ is relatively small, resulting in a relatively smaller $\tilde{f}(t_{obs})=8.2$. But, for ``m" model, the increase slope of $\gamma_{ch}$ is large (g=1.0), and the $x/x_c$ value decreases significantly with radius $r$, leading to a large $\tilde{f}(t_{obs})=97.3$. The above two factors finally give the calculated results.

The energy-evolving polarization is shown in Fig. 5 at $t_{obs}=0.5$ s and in Fig. 6 at $t_{obs}=2.5$ s. At an early time (e.g. $t_{obs}=0.5$ s), PDs of all the models increases with energy in the energy range from soft X-ray to MeV $\gamma-$ray band. From [2$b_i$] ([2$b_m$]) to [2$d_i$] ([2$d_m$]), PDs decrease slightly, whose reason was claimed above. The only difference of ``i" and ``m" models is $\gamma_{ch}$ evolution. There are two intersection points for the $\gamma_{ch}$ curves of ``i" and ``m" models: one is at $r_1=5.6\times10^{14}$ cm and the other is at $r_2=1.35\times10^{16}$ cm. The maximum radius on EATS at 0.5 s is $4.4\times10^{14}$ cm, which is smaller than $r_1$. Because the slope of $\gamma_{ch}$ curve of ``m" model is steeper, the flux ratio $\tilde{f}(0.5\ s)$ is higher for ``m" model\footnote{Taking $\nu_{obs}=1$ MeV as an example, $x$ is always larger than $x_c$ at $t_{obs}=0.5$ s for both models.}, leading to a larger PD. At late evolution times, e.g., $t_{obs}=2.5$ s, the evolutions of PDs are reversed, i.e., PDs of all calculated models, except for [2$b_i$],  roughly decrease toward a high energy band. PAs are constant within the calculated energy band. Independent of models, the profiles of energy-PD curves for a mixed magnetic field with an aligned ordered part are very similar to the corresponding curves with a purely aligned ordered magnetic field.

\section{Conclusions and Discussion}

Taking the models proposed by Uhm \& Zhang (2015, 2016) and Uhm, Zhang \& Racusin (2018), with which the observed spectral lags and peak energy $E_{p,obs}$ evolution patterns of the GRB prompt phase can be recovered, we have investigated the time-resolved and energy-resolved polarization evolution of the GRB prompt emission phase.

It is not found that polarization evolutions have direct correlations with the spectral lag or the $E_{p,obs}$ patterns. Evolution pattern of $E_{p,obs}$ is hard-to-soft for ``i" models, while it is intensity-tracking for ``m" models. PD curves of all these models, including three ``i" models and three ``m" models, roughly decay all the way with time. Furthermore, for the energy-resolved polarization, PDs increase with observational frequency for all the models at early observational times, while they decrease with observational frequency for all the models (except for [2$b_i$]) at late times. The PD evolution behavior seems to be unidentified for the hard-to-soft ``i" models and intensity-tracking ``m" models, so we do not find the direct correlations between the polarization properties and the $E_{p,obs}$ patterns. The spectral lag is negative for [2$d_m$] model, while it is positive for [2$b_m$] and [2$c_m$] models. PD curves of [2$d_m$] model are very similar to those of [2$b_m$] and [2$c_m$] models. Therefore, it seems that there is no correlations between the polarization and spectral lag. In Uhm, Zhang \& Racusin (2018), the evolution patterns of the electron Lorentz factor $\gamma_{ch}$ affect evolving patterns of $E_{p,obs}$ and the spectral lag significantly, while its effect on polarization properties is relatively weak, because it affects both flux and polarized flux simultaneously. Meanwhile, physical conditions, of which polarizations are sensitive, would not affect $E_{p,obs}$ evolution and spectral lag significantly.

In our calculation, EATS effect is considered. We find that at some times this effect is very important in shaping the PD curves, especially when the physical conditions of the emitting regions change violently at low-latitude region with local $\theta\Gamma<1$. In an EATS, with increasing radius $r$ (i.e., decrease of $\theta$), the Doppler boosting is enhanced, while the comoving specific intensity decreases due to a decaying magnetic field. If the jet is viewed on-axis, two key parameters $\tilde{f}(t_{obs})$ and $\Gamma\theta$ mainly determines the final PD of the jet emission. When $\theta<1/\Gamma$, the local emission is ``low-latitude" emission with high local PD; while when $\theta>1/\Gamma$, the local radiation is ``high-latitude" emission with lower local PD. Generally speaking, a relatively large $\tilde{f}(t_{obs})$ value will indicate a larger jet PD because the proportion of its low latitude emission (with high local PD) is higher. For the same $\tilde{f}(t_{obs})$ value, the larger range the $\Gamma\theta$ spans over an EATS, the smaller the jet PD is.

The time-evolving PD decays all the way during the GRB prompt phase. For the one-zone model discussed in this paper, several key parameters that affect polarization properties (e.g., the bulk Lorentz factor $\Gamma$, $\theta(r_{on})$ and $\theta(r_{off})$) may not only change with observational time $t_{obs}$, but also vary on one EATS of a fixed $t_{obs}$, which makes our analysis become very difficult. The energy-evolving PD at 0.5 s increases with energy, while it roughly decays toward high-energy band at 2.5 s. In Lundman et al. (2018), it is predicted that PD at MeV peak and above energy band are zero under the dissipative photosphere model, while our results show that MeV peak and the above high energy band can be highly polarized in the synchrotron model. Future energy-resolved polarization observations of the GRB prompt phase can distinguish between the two models.

Single-energy electrons are assumed in this paper, which follows the treatment of Uhm, Zhang \& Racusin (2018). The only difference of the polarization properties with single-energy electrons to power-law distributed electrons is their PD in an ordered magnetic field $\Pi_0$. For the single-energy electrons, $\Pi_0$ can range from $0.5$ to $1$ with $x$ ranging from $x\ll1$ to $x\gg1$, while it depends on the spectral index $p$ of power-law electrons (i.e., $\Pi_0=(p+1)/(p+7/3)$). For the magnetic reconnection model, PIC simulations give $p=4\sigma^{-0.3}$ (Sironi \& Spitkovsky 2014; Guo et al. 2015; Kagan et al. 2015; Werner et al. 2016), where $\sigma$ is the magnetization parameter. With the magnetic reconnection going on, $\sigma$ decays, leading to decrease of power-law index $p$. Thus, if the magnetic reconnection stops roughly at $\sigma\sim1$, $p$ reaches its maximum value of $4$ and $\Pi_0$ also reaches its maximum value of $\sim0.8$.

The most possible MFC in the magnetic reconnection models is a mixed magnetic field. Because the polarization degree in an ordered magnetic field gives the upper limit of that in a corresponding mixed magnetic field (Lan et al. 2019), they are also considered in this paper. Polarization properties of synchrotron emission in a mixed magnetic field with an aligned ordered part are very similar to those of a purely aligned ordered magnetic field, only with a smaller PD value. This is consistent with the assumptions we made for the mixed magnetic field. The random part of a mixed magnetic field is assumed to be isotropic in the 3-dimensional space and PD of such a random field is always zero. Polarization properties of such mixed magnetic field are mainly determined by its ordered part and the role of the random part gives an additional unpolarized flux to suppress the final PD. Here, we take the ratio of magnetic field strength of the ordered part to the random part (i.e., $\xi_B$) as a constant. For the magnetic dissipation model (e.g., ICMART model, Zhang \& Yan 2011), the $\xi_B$ parameter would decrease with time because the reconnection process destroys the ordered magnetic field. Because PD values decrease with a decaying $\xi_B$ (Lan et al. 2019), a decaying $\xi_B$ leads to a faster decaying PD curves than that of constant $\xi_B$."

\acknowledgements
We would like to thank the referee for his/her very careful and helpful comments and suggestions that have allowed us to improve the
presentation of this manuscript significantly. This work is supported by the National Key Research and Development Program of China (grant no. 2017YFA0402600) and the National Natural Science Foundation of China (grant no. 11833003 and 11903014). M.X.L is also supported by the Natural Science Foundation of Jiangsu Province (grant No. BK20171109) and the Fundamental Research Funds for the Central Universities, in part by National Science Foundation of China (NSFC) under Grant No. 11847310 and the Seeds Funding of Jilin University.

\begin{figure}
\includegraphics[angle=0,scale=0.6]{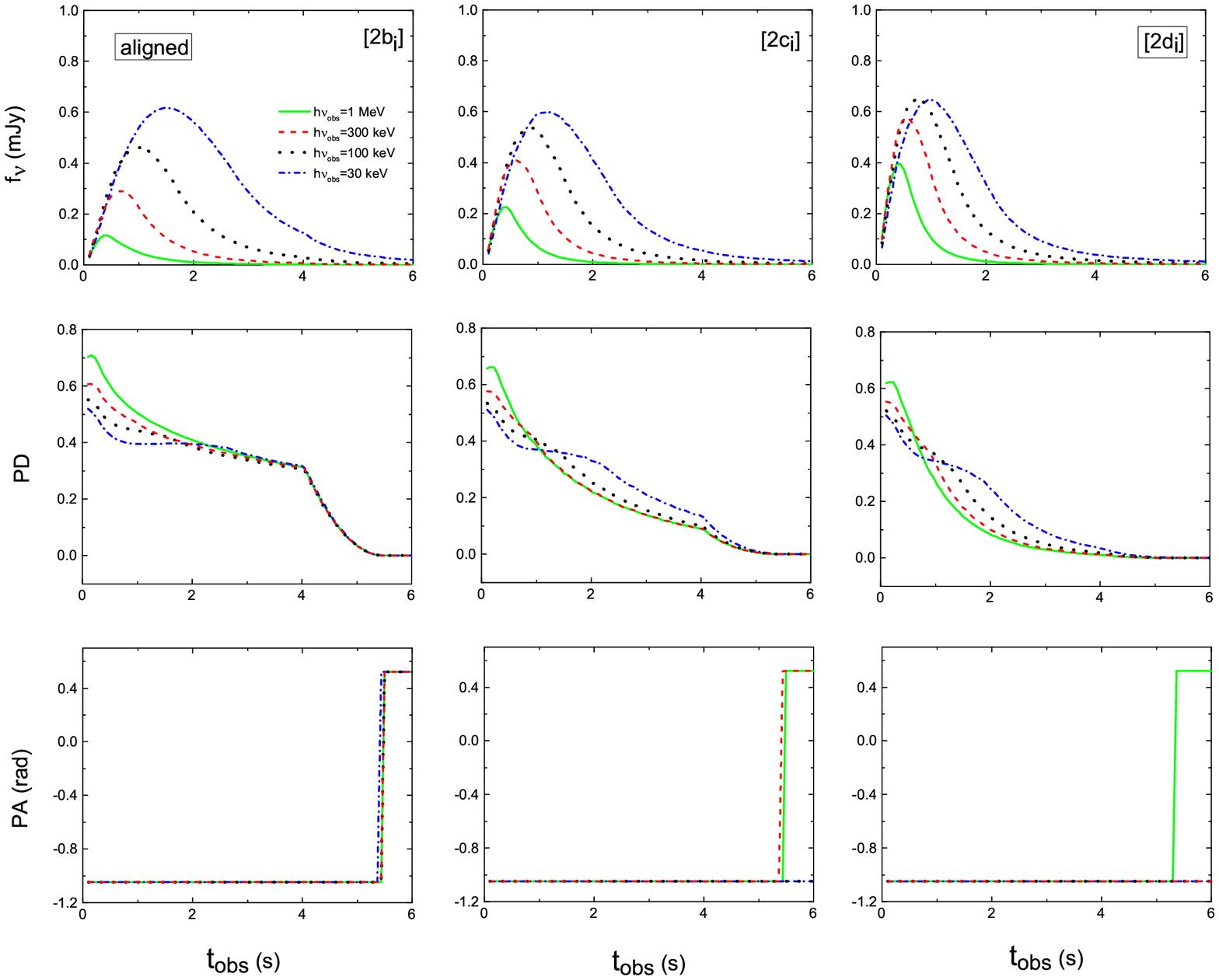}
\caption{Light curves and polarization evolution of three ``i" models with a purely ordered aligned magnetic field. The top, middle and bottom panels show the light curves, PD and PA curves, respectively. The green solid, red dashed, black dotted and blue dash-dot lines correspond to the observational photon energies of 1 MeV, 300 keV, 100 keV and 30 keV. The left, middle and right panels are corresponding to models [2$b_i$], [2$c_i$] and [2$d_i$], respectively.}  \label{fig1}
\end{figure}

\begin{figure}
\includegraphics[angle=0,scale=0.6]{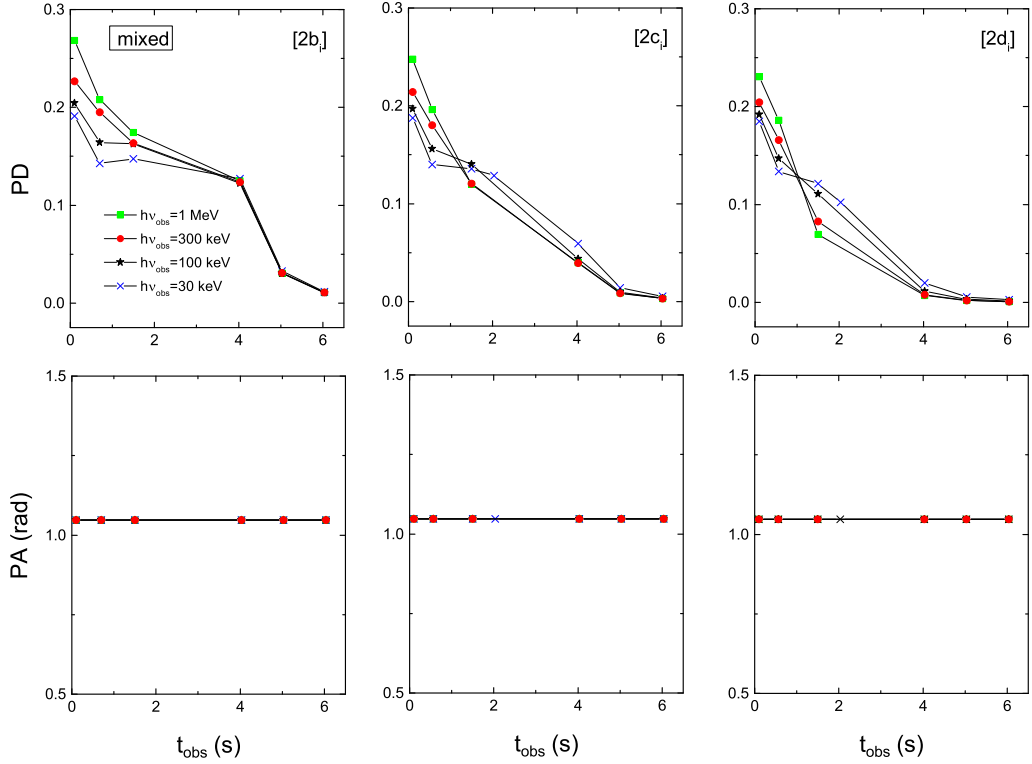}
\caption{Polarization properties for ``i" models for a mixed magnetic field with an aligned ordered part. The upper and lower panels exhibit the PD and PA curves, respectively. The green diamonds, red circles, black stars and blue crosses are our calculating points and correspond to 1 MeV, 300 keV, 100 keV and 30 keV, respectively. The left, middle and right panels are for model [2$b_i$], [2$c_i$] and [2$d_i$], respectively.} \label{fig1}
\end{figure}

\begin{figure}
\includegraphics[angle=0,scale=0.6]{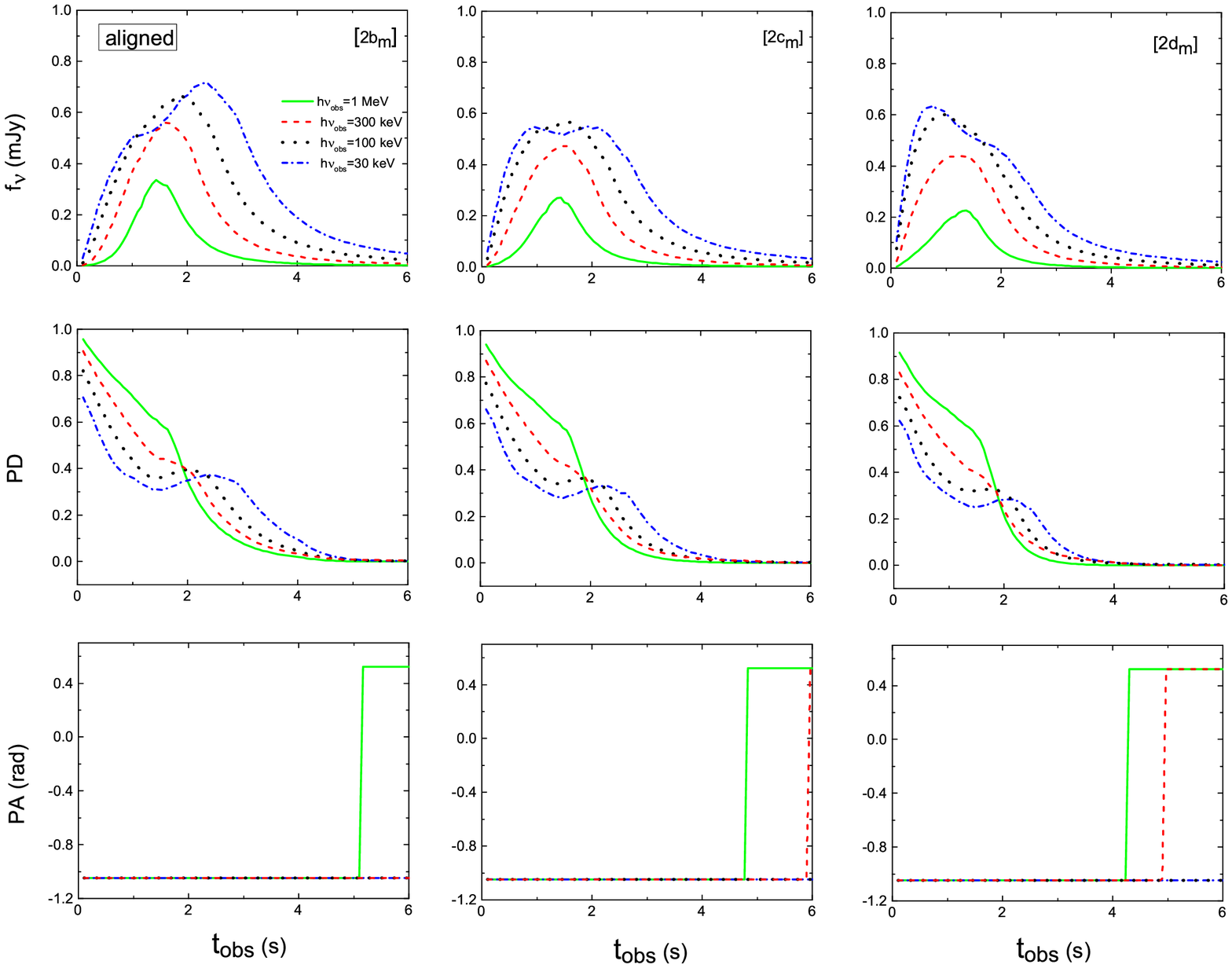}
\caption{Same as Fig. 1, but for model ``m".} \label{fig1}
\end{figure}

\begin{figure}
\includegraphics[angle=0,scale=0.6]{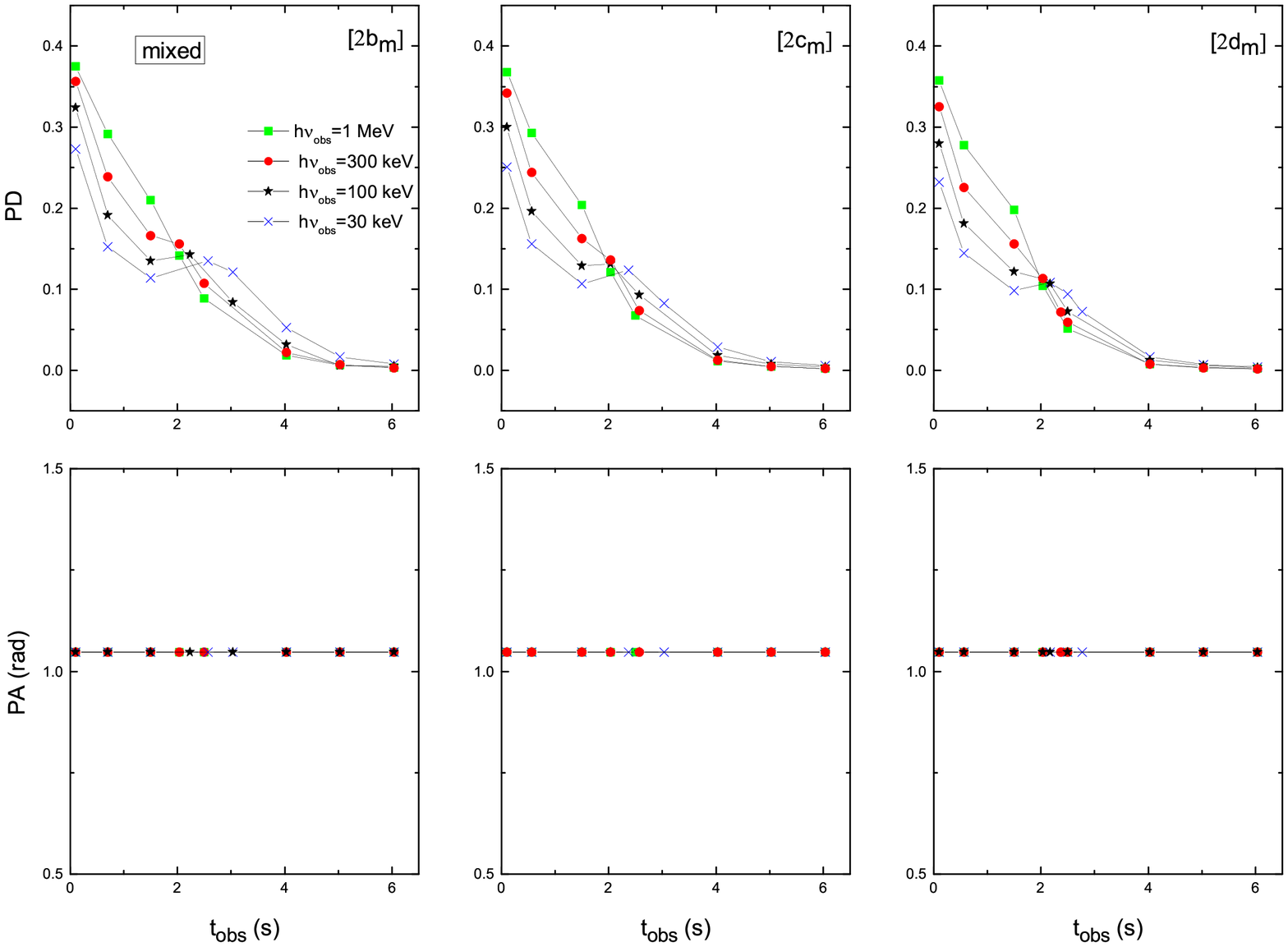}
\caption{Same as Fig. 2, but for model ``m".}  \label{fig1}
\end{figure}

\begin{figure}
\includegraphics[angle=0,scale=0.6]{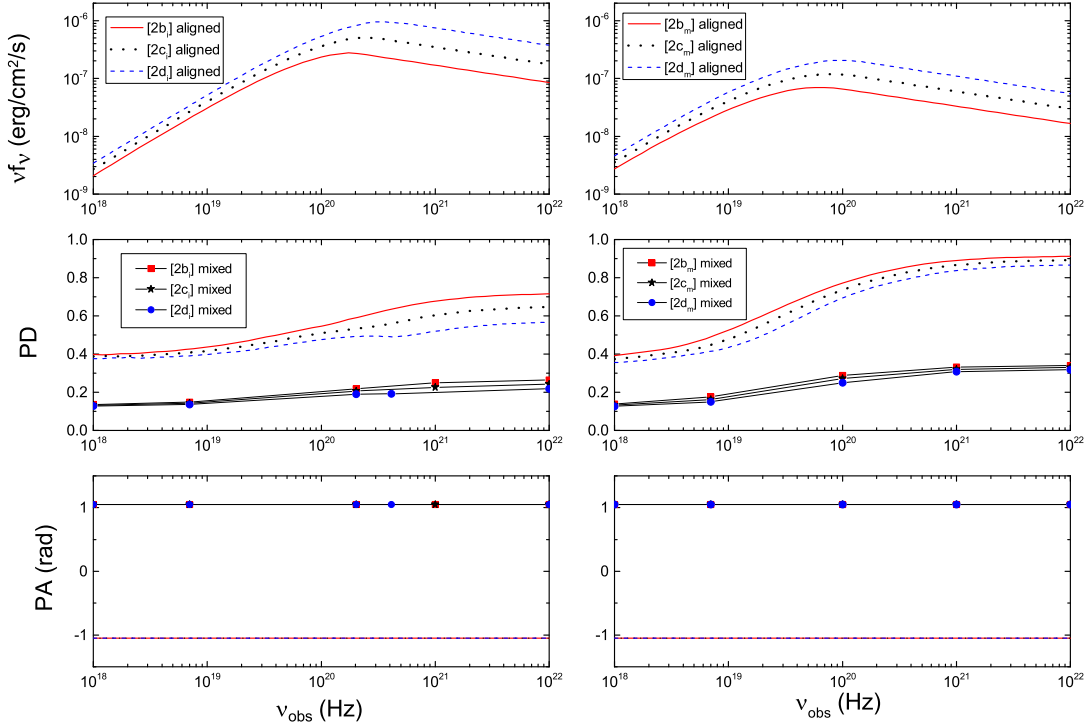}
\caption{Energy-evolved polarization properties with a purely ordered aligned magnetic field at $t_{obs}=0.5$ s. The top, middle and bottom panels show the energy spectrum, PD and PA. The left and right panels correspond to ``i" and ``m" models, respectively. The red solid, black dotted and blue dashed lines are for models [2$b_i$] ([2$b_m$]), [2$c_i$] ([2$c_m$]) and [2$d_i$] ([2$d_m$]). Red diamonds, black stars and blue circles are our calculating points of a mixed magnetic field with an aligned ordered part for [2$b_i$] ([2$b_m$]), [2$c_i$] ([2$c_m$]) and [2$d_i$] ([2$d_m$]) models. Because MFC affects a light curve slightly, we do not show the light curves with a mixed magnetic field.} \label{fig1}
\end{figure}

\begin{figure}
\includegraphics[angle=0,scale=0.6]{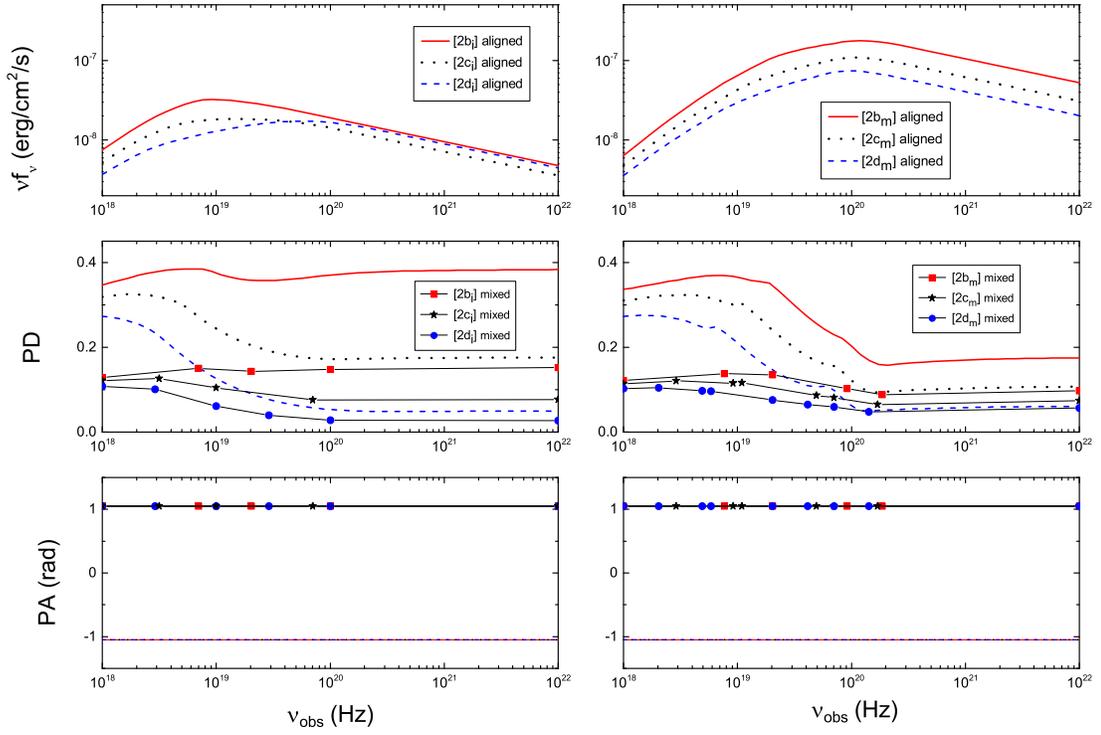}
\caption{Same as Fig. 5, but calculated at $t_{obs}=2.5$ s.}  \label{fig1}
\end{figure}

\end{document}